\documentclass[12pt,onecolumn,journal]{IEEEtran}
\usepackage{latexsym}
\usepackage{float}
\usepackage{amsfonts}
\usepackage{amsbsy}
\usepackage{amssymb}
\usepackage{times}
\usepackage{graphicx}
\usepackage{enumerate}
\usepackage[usenames]{color}
\usepackage[dvips]{pstcol}
\usepackage{epstopdf}
\usepackage{cite}
\usepackage{amssymb}
\usepackage{amsfonts}
\usepackage{graphicx}
\usepackage{epsfig}
\usepackage{psfrag}
\usepackage{xcolor}
\usepackage{amsfonts, bm}
\usepackage{epstopdf}
\usepackage{cite}
\usepackage{color}
\usepackage{xcolor}
\usepackage{subfig}
\usepackage{verbatim}
\usepackage{multirow}
\usepackage{array}
\usepackage{booktabs}
\usepackage{amsthm}
\usepackage{makecell}

\usepackage[linesnumbered, ruled]{algorithm2e}
\usepackage{algpseudocode}
\usepackage{amsmath}

\IEEEoverridecommandlockouts

\begin{document}
	
\title{Deep-Unfolding for Next-Generation Transceivers}

\author{ Qiyu Hu, Yunlong Cai, Guangyi Zhang, Guanding Yu, and Geoffrey Ye Li
\thanks{ Q. Hu, Y. Cai, G. Zhang, and G. Yu are with the College of Information Science and Electronic Engineering, Zhejiang University, Hangzhou 310027, China (e-mail: qiyhu@zju.edu.cn; ylcai@zju.edu.cn; zhangguangyi@zju.edu.cn; yuguanding@zju.edu.cn). 

G. Y. Li is with the Department of Electrical and Electronic Engineering, Imperial College London, London SW7 2AZ, UK (e-mail: geoffrey.li@imperial.ac.uk).  } }

\maketitle
\vspace{-3.3em}
\begin{abstract}

The stringent performance requirements of future wireless networks, such as ultra-high data rates, extremely high reliability and low latency, are spurring worldwide studies on defining the next-generation multiple-input multiple-output (MIMO) transceivers.
For the design of advanced transceivers in wireless communications, optimization approaches often leading to iterative algorithms have achieved great success for MIMO transceivers. However, these algorithms generally require a large number of iterations to converge, which entails considerable computational complexity and often requires fine-tuning of various parameters. 
With the development of deep learning, approximating the iterative algorithms with deep neural networks (DNNs) can significantly reduce the computational time. 
However, DNNs typically lead to black-box solvers, which requires amounts of data and extensive training time. To further overcome these challenges, deep-unfolding has emerged which incorporates the benefits of both deep learning and iterative algorithms, by unfolding the iterative algorithm into a layer-wise structure analogous to DNNs.
In this article, we first go through the framework of deep-unfolding for transceiver design with matrix parameters and its recent advancements. Then, some endeavors in applying deep-unfolding approaches in next-generation advanced transceiver design are presented. Moreover, some open issues for future research are highlighted.  
\end{abstract}


\IEEEpeerreviewmaketitle

\section{Introduction}
Sixth generation (6G) wireless systems will embrace a horizon of new Internet-of-Everything applications, such as autonomous vehicles,  industrial automation, virtual reality (VR), and more. To accommodate these applications, future 6G networks will need to meet much more stringent performance requirements than the fifth generation (5G), such as ultra-high data rates and energy efficiency, extremely high reliability and low latency, as well as global coverage and connectivity. These demanding performance requirements are hard to be achieved by current technologies, thus requiring the development of revolutionary ones \cite{6G}.

Advanced transceiver design for next-generation multiple-input multiple-output (MIMO) systems is a core technology that comprises innovative designs in transmit and receive processing architecture, beamforming, detection, channel estimation and feedback. It has an ambitious goal  of boosting the future 6G network performance in broader Terahertz frequency bands, diverse applications, and new systems and networks.
To satisfy these requirements, iterative optimization algorithms have been considered for future transceivers. Although these algorithms exhibit satisfactory system performance in many settings, there are still many obstacles to their applications in practical communication systems. In particular, they require knowledge of various channel parameters and fine-tuning of regularization parameters. In addition, a large number of iterations cause high computational complexity and make the deployment difficult.

Data-driven deep learning approaches can be used to replace these iterative algorithms \cite{qin_phydl}. These approaches focus on treating the algorithm as a ``black-box" and learning the mapping between the input and output with deep neural networks (DNNs). 
Despite acceptable performance offered by these algorithms, they suffer from poor interpretability and generalization ability, and their performance is hard to guarantee in practical scenarios with dynamic environments. 
Furthermore, it highly relies on the training of a huge amount of labeled data and requires a long training time. However, for practical communication devices, both data and computational resources are scarce, thus these algorithms need to be further improved to overcome the aforementioned shortcomings.

To address these issues, deep-unfolding neural network (DUNN), which can be regarded as a model-driven deep learning technique \cite{ModelPHY}, appears to be a promising solution, where the iterations are unfolded into a layer-wise structure and a set of trainable parameters are introduced to improve system performance. The DUNNs are shown to outperform the black-box DNNs and can approach or even surpass the performance of iterative algorithms with much reduced computational complexity, satisfactory interpretability, and generalization ability \cite{AlgorithmUnroll}. 

\begin{figure}[t]
\begin{centering}
\includegraphics[width=0.6\textwidth]{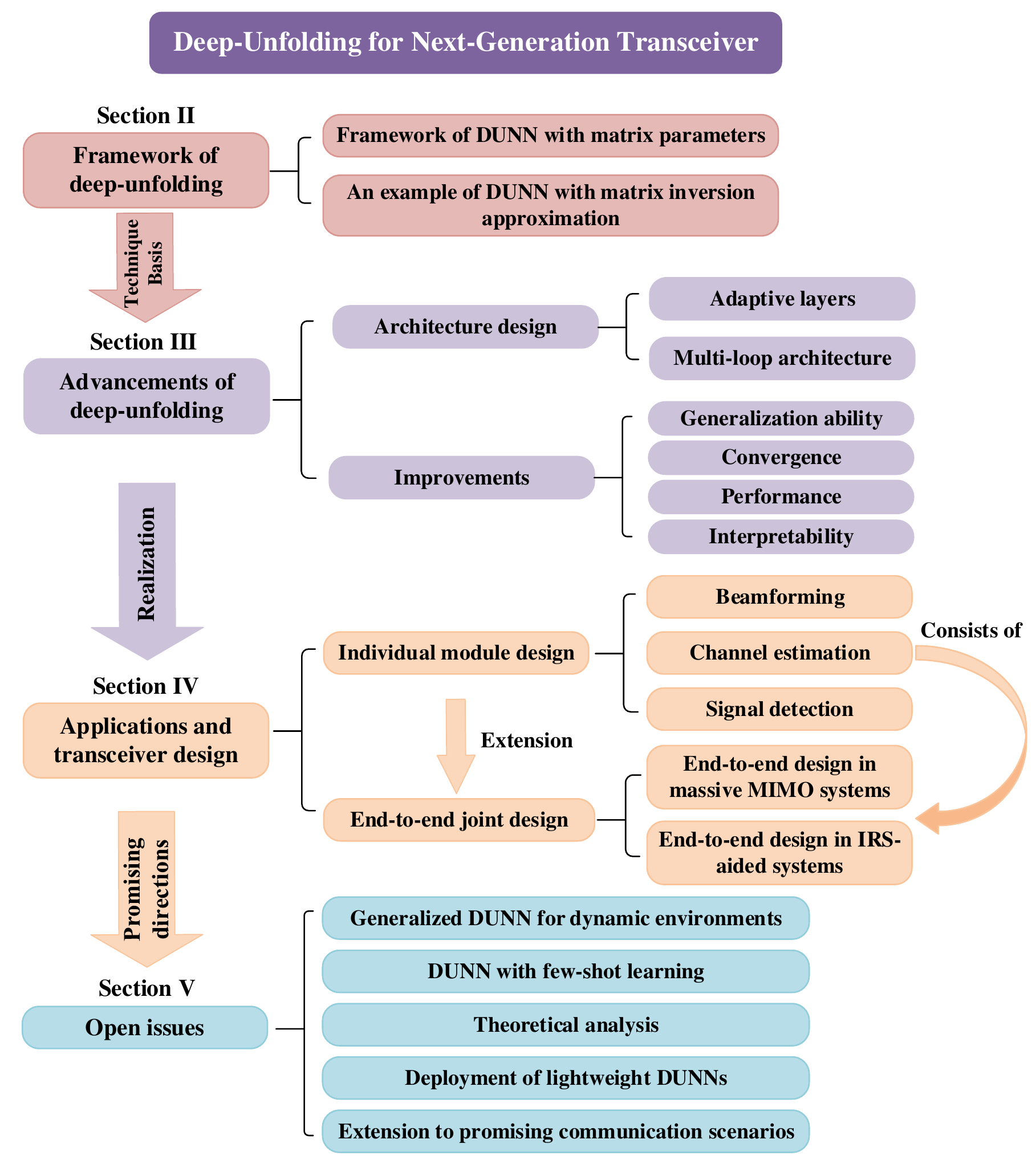}
\par\end{centering}
\caption{The organization and contribution of this article.}
\label{OrgaPaper}
\end{figure}

This article provides a comprehensive overview of the latest developments in DUNNs in transceiver design for next-generation MIMO systems. The detailed structure of this article is shown in Fig. \ref{OrgaPaper}. In particular, 
Section \ref{FrameworkDUNN} introduces a general framework of DUNN for transceiver design in wireless communications with complex matrix parameters, and the generalized chain rule (GCR) for deriving the gradients to better train the DUNN. Based on this framework, in Section \ref{AdvancementsDUNN}, we present the advancements of the DUNN architecture, such as the adaptive depth, the multi-loop structure, and the improvement of generalization ability. 
Based on these advancements, in Section \ref{DUNNTransceiver}, we present some application examples of deep-unfolding enabled MIMO transceivers, such as beamforming and channel acquisition, and propose to jointly design these modules of transceiver in an end-to-end manner. Finally, we discuss some open issues in the area of DUNN-enabled next-generation transceivers in Section \ref{OpenIssue}. 

\section{Framework of Deep-Unfolding} \label{FrameworkDUNN}
In this section, we first introduce the framework of DUNN for transceivers with matrix parameters and then design its structure.

\subsection{Framework of DUNN with Matrix Parameters}

The design of a typical DUNN for wireless transceivers includes three components: problem itself, iterative optimization algorithm, and training data. The idea of DUNN is to unfold an iterative optimization algorithm into a multi-layer structure, similar to DNNs \cite{qiyu_unfoldwmmse}. 
In particular, the objective function in the problem can be regarded as the loss function of DUNN and its outputs are the variables to be optimized in the problem. Moreover, a number of trainable parameters are introduced to improve the system performance and approximate the high-complexity operations. The training data aims to help the DUNN find the satisfactory trainable parameters by training with the stochastic gradient descent (SGD) method.
In brief, by integrating the trainable parameters and iterative algorithm, an approximation of the iterative algorithm is provided to address the problem with lower computational complexity. 
As deep-unfolding is developed by taking into account both the problem structure and the corresponding iterative algorithm, it generally requires less training data and is more effective to be trained with less training time, compared with the black-box DNNs.

A general framework for deep-unfolding is initially conceived in \cite{qiyu_unfoldwmmse}. 
The proposed framework is formulated in the matrix form to better fit the applications for beamforming design in massive MIMO systems, which consists of two stages. 
In the forward propagation, one iteration of the iterative algorithm is designed as a layer of DUNN and some trainable matrix parameters are introduced to approximate the high-dimensional matrix inversion operation.
In the back-propagation, a closed-form gradient expression is derived based on the proposed GCR in matrix form. Then, the SGD is employed to train these parameters. Note that most of DNNs are implemented in widely-used platforms, e.g., Pytorch, and the gradients are computed using the automatic gradient mechanism. However, the gradient of some operations cannot be computed in the platform, e.g., the inversion and determinant of complex matrices.
In comparison, the derived GCR can compute the closed-form gradients of all operations, which are more accurate and more efficient. This is because unnecessary computational graphs can be escaped when performing back propagation, i.e., the gradients of the unnecessary parameters can be avoided.

\subsection{Some Examples of DUNN}

\begin{figure*}[t]
\begin{centering}
\includegraphics[width=0.89\textwidth]{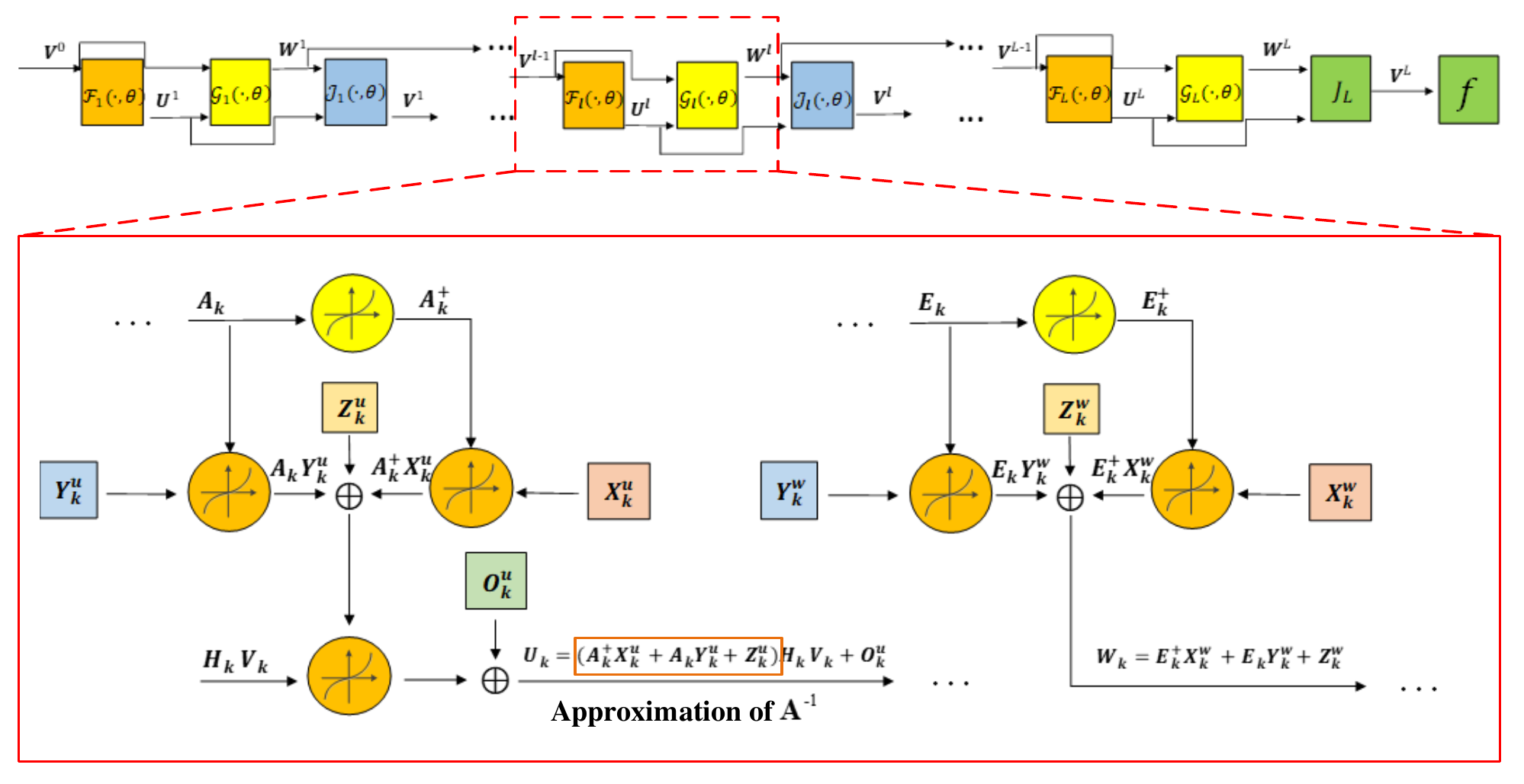}
\par\end{centering}
\caption{The architecture of iterative WMMSE algorithm-driven DUNN, where $\mathcal{F}$, $\mathcal{G}$, and $\mathcal{J}$ denote the layers of DUNN for updating different variables.}
\label{UW_frame}
\end{figure*}

To take full advantage of the structure of iterative algorithms, most parts of the algorithms are reserved and the trainable parameters are introduced to approximate the high-complexity operations. A typical example from \cite{qiyu_unfoldwmmse} is shown in Fig. \ref{UW_frame}, which unfolds the well-known iterative weighted minimum mean-square error (WMMSE) algorithm into a multi-layer structure for beamforming design in multiuser multiple-input multiple-output (MU-MIMO) systems, to solve the sum-rate maximization problem. It is called iterative algorithm induced deep-unfolding neural network (IAIDNN), where the matrix inversion operation in WMMSE is approximated by the matrix multiplications of trainable parameters. 
Particularly, inspired by the first-order Taylor expansion of the inverse matrix, $\mathbf{A}^{-1}$ can be approximated by $\mathbf{A}^{\dagger}\mathbf{X} + \mathbf{A}\mathbf{Y} + \mathbf{Z}$, where $\mathbf{X}$, $\mathbf{Y}$, and $\mathbf{Z}$ are introduced trainable parameters \cite{qiyu_beamselect}. The proposed non-linear function, $\mathbf{A}^{\dagger}$, denotes taking the reciprocal of each element in the diagonal elements of $\mathbf{A}$ and setting the non-diagonal elements to $0$, which is an accurate approximation for the matrix inversion dominated by diagonal elements. The success of this approximation shows that the performance of DUNN can be improved by combining the knowledge about matrix inversion with deep learning.

Moreover, the DUNN can also be integrated with the conventional black-box DNN. In \cite{kkai_mixedtimesclae}, the black-box CRNet is integrated into the DUNN to replace high-complexity operations in channel state information (CSI) feedback. Aided by these black-box DNNs with powerful capabilities of learning and feature extraction, DUNNs can burst considerable potential in wireless communications.

\section{Why Deep-Unfolding?} \label{AdvancementsDUNN}

\begin{figure}[t]
\begin{centering}
\includegraphics[width=0.72\textwidth]{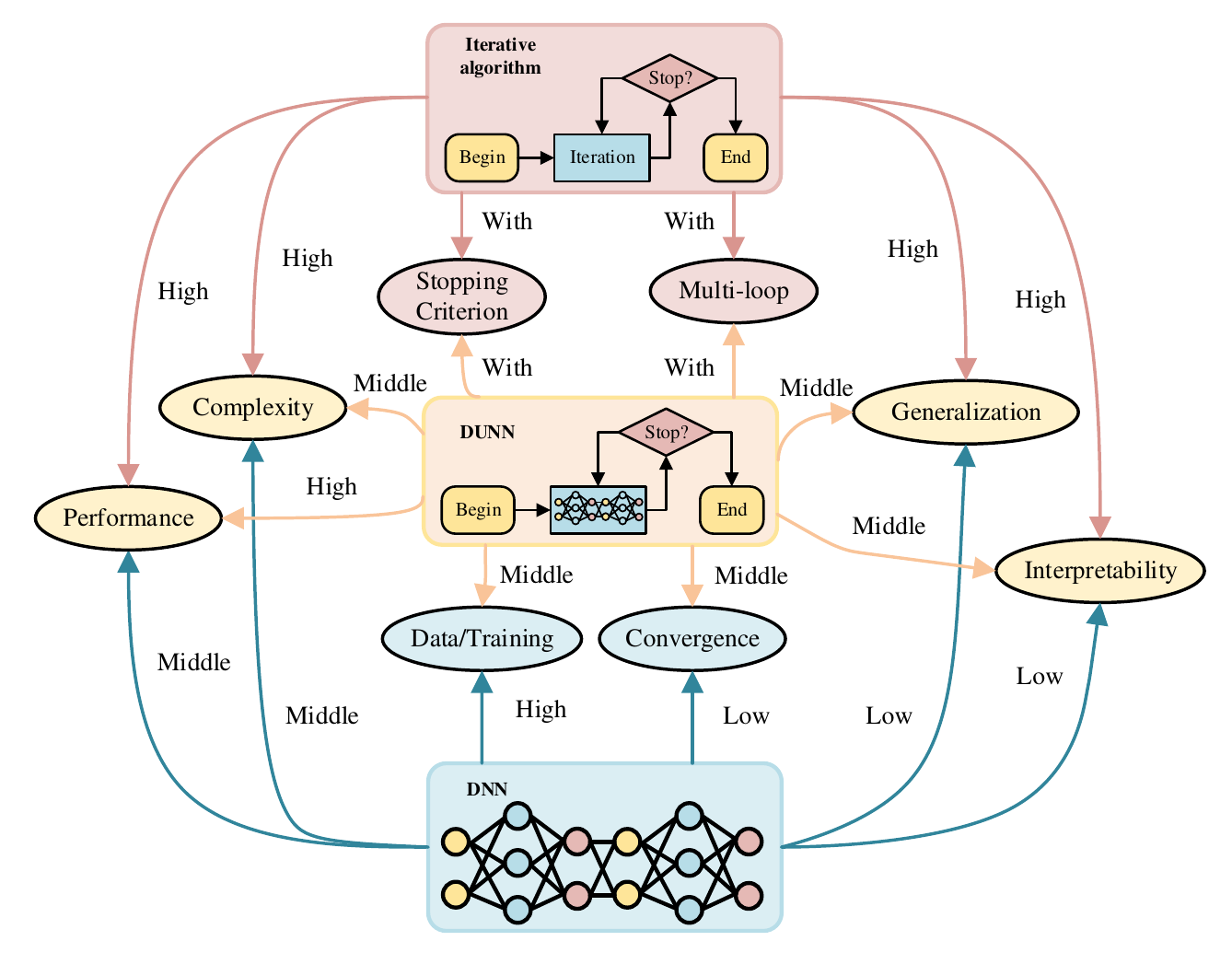}
\par\end{centering}
\caption{Comparison and relation among the iterative algorithm, black-box DNN, and DUNN.}
\label{CompRela}
\end{figure}

In this section, efforts have been made to design some advancements of the DUNN architecture, such as adaptive depth and multi-loop structure, etc., to improve its generalization ability, convergence, performance, and interpretability. Fig. \ref{CompRela} shows some characteristics of the iterative algorithm, black-box DNN, and DUNN with advanced architecture.

\subsection{Adaptive Depth}
\begin{figure*}[t]
\begin{centering}
\includegraphics[width=0.8\textwidth]{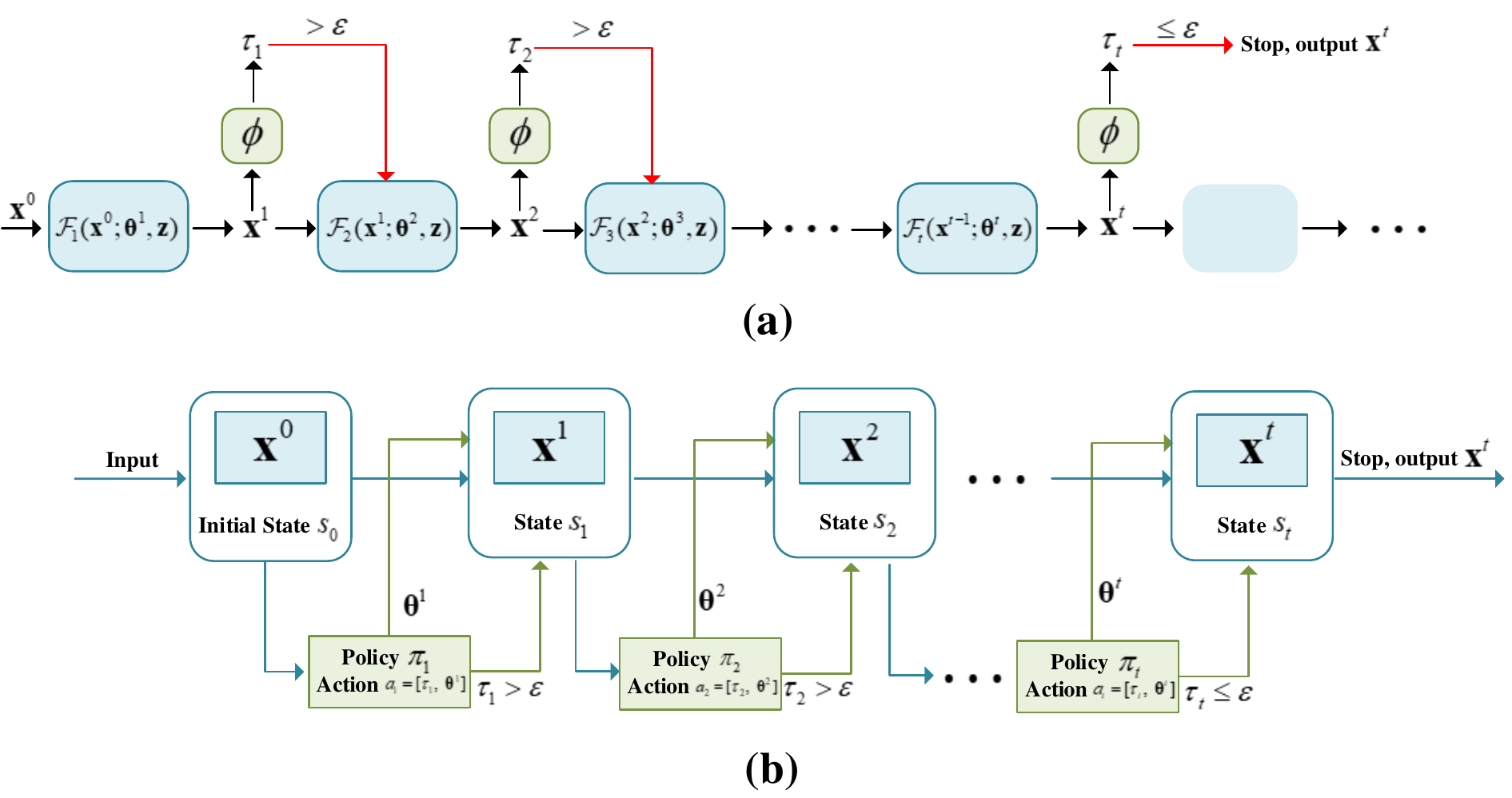}
\par\end{centering}
\caption{(a) DUNN with adaptive depth; (b) DDPG-driven deep-unfolding framework with adaptive depth.}
\label{StopPredict}
\end{figure*}

Most of the iterative algorithms can adjust the number of iterations for different inputs by setting stopping criterion and comparing whether the results after each iteration satisfy the conditions. In comparison, the depth of DUNN, i.e., the number of layers, is pre-determined and fixed. However, the optimal number of layers required for convergence varies from different inputs. In addition, the computational complexity of the algorithm depends on the depth. Moreover, in real communication scenarios, the characteristics of input samples may change frequently and the dynamic DUNN with adaptive depth is more practical and effective \cite{qiyu_adaptivedepth}. 
The DUNN with adaptive depth in \cite{chenwei_depth} can solve the signal recovery problem, where the depth for convergence changes with the sparsity of the signal vector. As shown in Fig. \ref{StopPredict}(a), $\mathcal{F}(\cdot)$ denotes the mapping of each unfolding layer, function $\phi$ outputs $\tau$ and the DUNN stops to output the results when $\tau>\varepsilon$, where $\phi$ is a part of DUNN and is trained together with DUNN.

To this end, for the framework of deep deterministic policy gradient (DDPG)-driven DUNN with adaptive depth in \cite{qiyu_adaptivedepth}, the trainable parameters are learned by DDPG, rather than updated by the SGD, to avoid gradient explosion. As in Fig. \ref{StopPredict}(b), the optimization variables, trainable parameters, and architecture of DUNN are designed as the state, action, and state transition of DDPG, respectively. To achieve the adaptive depth, a halting score scheme is created to indicate when to stop. Then, this framework is employed to unfold the sparse Bayesian learning (SBL) algorithm to deal with the channel estimation problem in massive MIMO systems. The success of \cite{qiyu_adaptivedepth, chenwei_depth} inspires us that the efficiency of DUNN can be further improved and its architecture can be better designed. 

\subsection{Multi-Loop Structure}
The iterative algorithms can be divided into different kinds, according to the number of iteration loops.  
For example, the WMMSE is a single-loop algorithm while the penalty dual decomposition (PDD) algorithm is a multi-loop algorithm \cite{gy_pdd}. Compared with single-loop algorithms, multi-loop algorithms are generally employed to solve more complicated problems  involving complex objective functions and coupled constraints. 
Besides, the total number of iterations of a multi-loop algorithm are extremely large. Therefore, it is challenging to design a DUNN based on a multi-loop algorithm, which involves a larger number of layers and is difficult to train. 

The first attempt has been made in \cite{gy_pdd} to design a multi-loop DUNN, which solves the spectrum efficiency maximization problem in massive MIMO systems with a hybrid beamforming architecture. It unfolds the PDD algorithm into a multi-loop architecture consisting of layers in the inner and outer loops. Moreover, the unconstrained iterative gradient descent (GD) hybrid beamforming algorithm in \cite{shuhan_symbol} can achieve the minimum symbol-error rate (MSER) through the application of kernel density estimation and a parametric representation of the unit-modulus matrix entries. Then, the iterative GD algorithm is unfolded into a multi-loop structure wherein trainable parameters are introduced to accelerate the convergence and improve the MSER performance. 
It has been proved that the multi-loop DUNN scheme can find a stationary solution by making full use of the structure of the multi-loop iterative algorithm and introduced trainable parameters \cite{shuhan_symbol}.

\subsection{Generalization Ability}
\label{General}

To make the DUNNs employable in practical systems, the generalization ability is a key issue. Although DUNN is designed based on an iterative algorithm, the introduced trainable parameters inspired by black-box DNNs may severely impact its generalization ability. The generalization ability of DUNNs is mainly analyzed in \cite{qiyu_unfoldwmmse}, \cite{qiyu_adaptivedepth}, and \cite{chenwei_depth} from the following three aspects: 
\begin{itemize}
\item  The generalization ability to different SNRs: For a DUNN trained under a certain SNR,  what is the performance loss when it is directly employed in other SNR scenarios?

\item  The generalization ability of system configuration mismatch: How can we generalize the DUNN to a system with different configurations when there is a system configuration mismatch between the training phase and testing phase, such as the number of antennas, users, and radio frequency (RF) chains? Note that in this case the testing data cannot be directly input into the DUNN since there exists a data dimensionality mismatch.

\item  The generalization ability for various channel statistics: Generally, the channel samples under a certain statistic are the input of a DUNN, and training DUNN with these samples can achieve high performance under this statistic. However, if the channel statistics change, how can DUNN be adapted to this change and still achieve satisfactory performance? 
\end{itemize}	

From \cite{qiyu_unfoldwmmse}, \cite{qiyu_adaptivedepth}, and \cite{chenwei_depth}, the DUNN generalizes satisfactorily to mismatches in SNR, channel variety, and system configuration. In particular, the mismatch in SNR can be handled by involving the samples with a wide range of SNRs in the training procedure. Moreover, the mismatch in system configuration can be resolved by padding the corresponding input and trainable parameters of DUNN with zero values. In addition, transfer learning can be employed to improve the robustness of DUNN to the change of channel statistics.

\begin{figure}[t]
\begin{centering}
\includegraphics[width=0.57\textwidth]{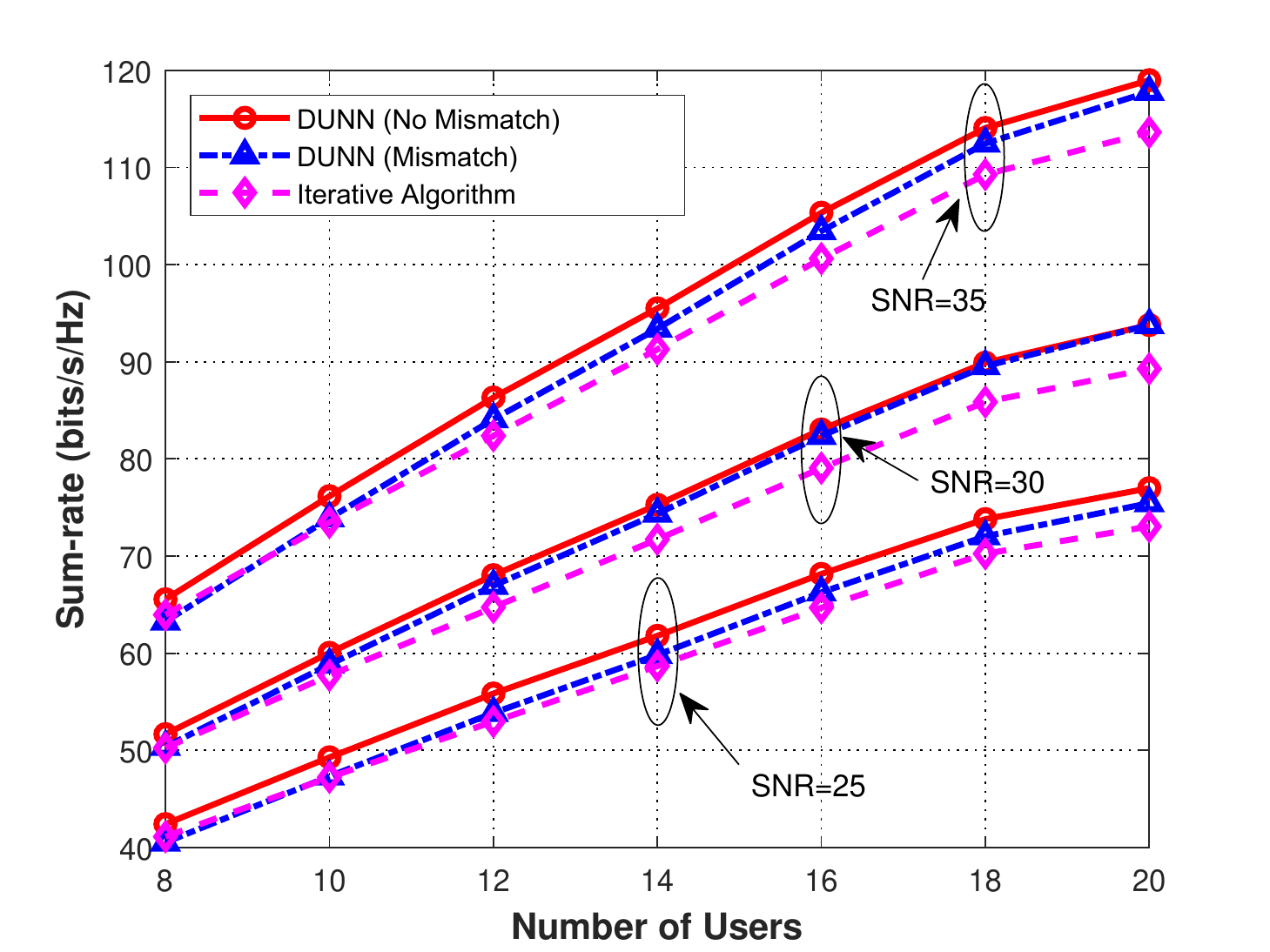}
\par\end{centering}
\caption{The generalization ability of DUNN.}
\label{MismatchK}
\end{figure}

Fig. \ref{MismatchK} presents the generalization ability of DUNN proposed in \cite{qiyu_beamselect} with the mismatch of SNR and the number of users, $K$. The model is trained at $K=20$ and SNR $=30$ dB and tested at different values of SNR and $K$. It can be readily seen that the mismatched DUNN outperforms the iterative algorithm. The small performance loss between the DUNN with mismatch and the DUNN without mismatch demonstrates its satisfactory generalization ability. 

\subsection{Training Strategies, Convergence, and Interpretability }
Despite being induced by iterative algorithms, the training of DUNN is still challenging. An important issue is that both computing resources and training samples are costly in practical communication systems. Thus, the DUNN is expected to converge fast with a small number of training samples. Moreover, the transmission overhead and computational complexity should be reduced. To solve these problems, a mixed-timescale training method has been developed in \cite{kkai_mixedtimesclae}.
Another important issue is vanishing or exploding gradients, i.e., the gradient is prone to be zero or infinity in the back propagation, which seriously affects the update of trainable parameters. To address this issue, a training strategy for DRL-based DUNN has been proposed in \cite{qiyu_adaptivedepth}.

Note that the training settings of DUNN, i.e., learning rate, batch size, and optimizer, etc., have significant impacts on the convergence efficiency. Based on the numerical results in \cite{qiyu_unfoldwmmse} and \cite{qiyu_adaptivedepth}, a larger batch size generally leads to slower but more stable convergence performance, a smaller learning rate achieves higher system performance, and a larger learning rate leads to faster convergence speed. Therefore, an adaptive scheme can be designed to dynamically adjust the learning rate, to achieve a high system performance and fast  convergence speed. 

Interpretability plays an important role in designing communication systems, which theoretically guarantees system performance against uncertainties in wireless environments. Inspired by the analysis in iterative algorithms, the authors in \cite{qiyu_adaptivedepth,shuhan_symbol} demonstrate that one layer of DUNN can approach the performance of multiple iterations of the optimization algorithm, which indicates that the DUNNs achieve the performance close to the iterative algorithm with a greatly reduced number of layers.

\begin{figure*}[t]
\begin{centering}
\includegraphics[width=0.85\textwidth]{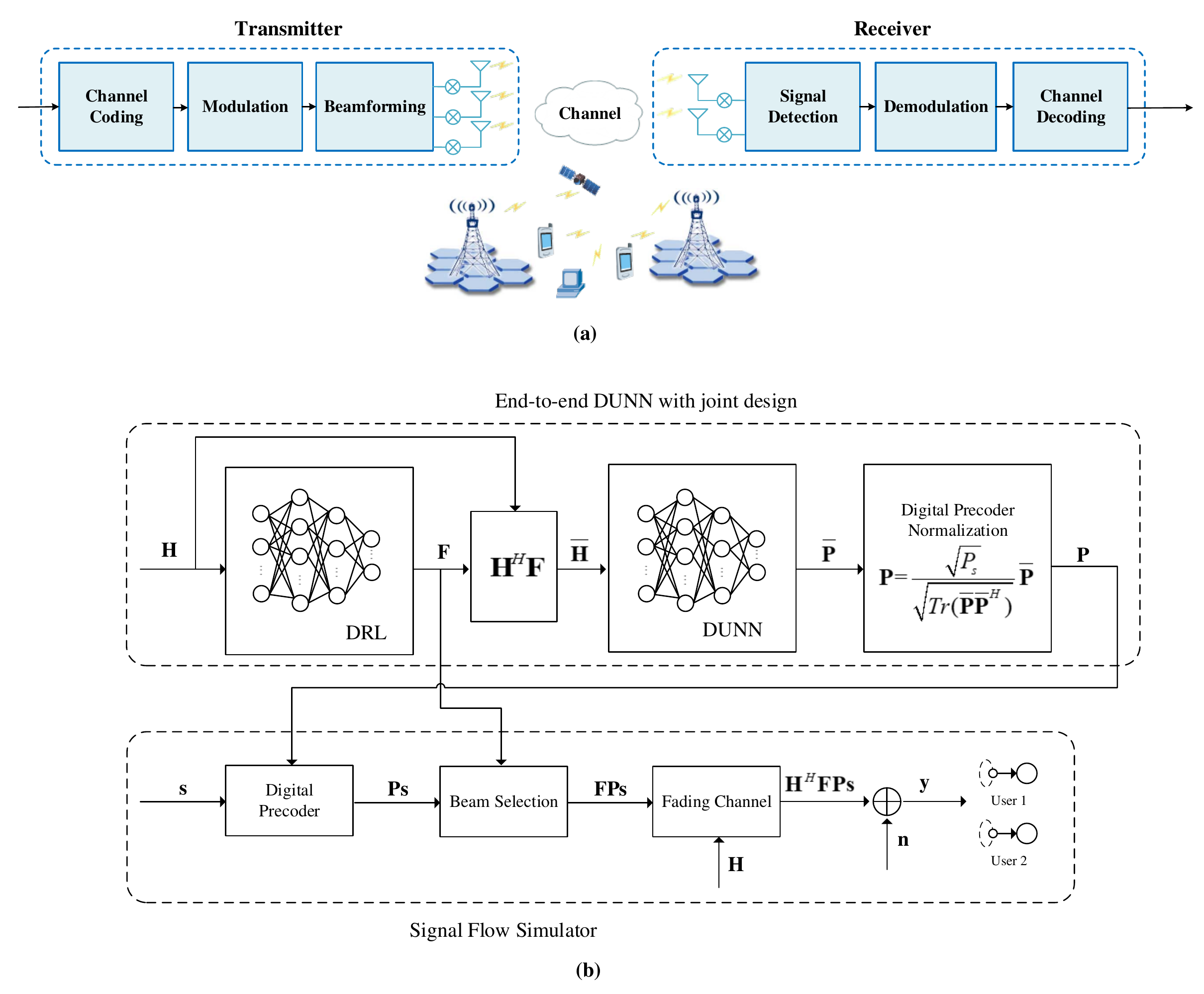}
\par\end{centering}
\caption{The physical-layer MIMO transceiver: (a) Overall architecture; (b) An example of end-to-end DUNNs for MIMO transceiver.}
\label{Transceiver}
\end{figure*}

\section{Deep-Unfolding Enabled MIMO Transceiver Technologies} \label{DUNNTransceiver}

In this section, the applications of DUNN in some key modules of MIMO transceiver design are illustrated, including beamforming, channel estimation, and signal detection. Then, we introduce some end-to-end DUNNs for joint transceiver design.

\subsection{Challenges of Future Physical-Layer Transceiver Design}
Physical-layer MIMO transceiver plays a vital role in next-generation communication systems, as shown in Fig. \ref{Transceiver}(a). In particular, the input bit streams successively pass through the channel coding, modulation, and beamforming modules. Then, the processed signal will propagate through wireless channels and be detected by the receiver. The receiver aims to recover the information bits, which consists of the signal detection, demodulation, and channel decoding modules, etc. 

As for the 6G wireless communications, it is of great importance to further scale up the number of antennas and expand the frequency band to meet the increasing traffic demands. Though the optimization-based transceiver has achieved considerable success, the newly raised demands intuitively cast challenges in practical implementation and deployment. Particularly, the increasing numbers of antennas and radio frequency (RF) chains will bring several difficulties in system design: (i) The extremely huge number of RF chains increase the difficulty of beamformer design and the hardware cost is high; (ii) The substantial antennas lead to a huge dimension of the CSI matrix, which seriously increases computational complexity of the channel estimation, CSI feedback, and transmission overhead; (iii) The complexity of traditional detectors, such as maximum likelihood detection, increases exponentially with the number of antennas. 
Therefore, efficient signal processing algorithms and transceiver design are required to achieve excellent performance with moderate computational complexity. Driven by deep-unfolding, the next-generation MIMO transceiver can be implemented by revising the modules with optimization algorithms in the conventional communication systems. In the following, we go through a series of applications of DUNN for transceiver design.

\subsection{Beamforming}
Among various new technologies of transceiver, beamforming, especially the hybrid beamforming has attracted great attention. In particular, the efficient DUNN in \cite{qiyu_unfoldwmmse} addresses the fully-digital beamforming for sum-rate maximization problem. Specifically, the classic iterative WMMSE algorithm is unfolded into a layer-wise structure, where a number of trainable parameters are introduced to replace the high-complexity operations. The DUNN in \cite{kkai_mixedtimesclae} achieves approaching performance to fully-digital beamforming with reduced complexity of hardware and computation. Moreover, the joint design of beam selection and digital beamforming matrices for mmWave MU-MIMO systems in \cite{qiyu_beamselect} maximizes the system sum rate. An efficient joint design framework is developed, which consists of a deep reinforcement learning (DRL)-based DNN and a DUNN to optimize the beam selection and digital beamforming matrices, respectively.

\subsection{Channel Estimation}
There are many effective channel estimation methods exploiting the sparsity of massive MIMO channels. Nevertheless, they generally have high computational complexity and some \emph{a priori} parameters are difficult to determine.  
The DUNN based on SBL algorithm in \cite{qiyu_adaptivedepth} tackles this problem. Specifically, a set of trainable parameters are introduced to improve the performance, which can be divided into two categories: (i) the existing parameters in the SBL-based algorithm, such as a \emph{a priori} parameters that are difficult to determine; (ii) the introduced trainable parameters to replace the operations with high computational complexity.

\subsection{Signal Detection}
The benefits of MIMO technology depend on efficient signal detection algorithms. Based on approximate message passing (AMP) and expectation propagation, a set of iterative algorithms have been developed for MIMO detection. 
These algorithms presume that the channels obey a specific distribution and are hard to adapt to different channel environments. To address this issue, the DUNN-driven detector in \cite{He_mimodetection}, named OAMP-Net2, is developed by unfolding the OAMP detector into a layer-wise structure. 
Specifically, the introduced trainable parameters are optimized to adapt to various channel environments and account for channel estimation errors. As a result, OAMP-Net2 is robust to a wide range of SNRs and can effectively handle channel correlation. 

\subsection{End-to-End Transceiver Design in Massive MIMO Systems}
Most DUNNs are proposed for designing individual modules in a transceiver, such as channel estimation, signal detection, and decoding with different optimization metrics. However, a transceiver is usually composed of several highly correlated modules. Thus, there exists a demand for joint designs, which  generally provide better performance than separate designs. However, the joint design faces more complicated problems with coupled constraints \cite{qiyu_beamselect}, which is difficult to address. 

Therefore, it is challenging to accomplish such joint design by simply employing data-driven black-box DNNs. Compared with designing the modules jointly with black-box DNNs, DUNNs are more interpretable and effective. 
Fig. \ref{Transceiver}(b) presents the end-to-end deep-unfolding framework for the joint design  proposed in \cite{qiyu_beamselect}, where the DRL can be replaced by other structures of DNN, e.g., CNN and DUNN, etc., to solve different problems. The signal flow simulates the process of signal transmission at the physical layer.   
To jointly train the DRL and DUNN in Fig. \ref{Transceiver}(b) in an end-to-end manner, a novel training method is proposed in \cite{qiyu_beamselect}. It reveals that the DUNN can be combined with other deep learning methods, allowing us to jointly design the modules of communication systems.

Inspired by \cite{qiyu_beamselect}, a novel end-to-end DUNN is developed to handle the sum-rate maximization problem in massive MIMO systems, which considers both channel estimation and hybrid beamforming \cite{kkai_mixedtimesclae}. 
In particular, two DUNNs are designed for channel estimation and hybrid beamforming, and the channel quantization and feedback are implemented with black-box DNNs. It achieves better performance than the separately designed DUNNs and black-box DNNs, which shows that the end-to-end learning with DUNNs and black-box DNNs is an effective approach for the design of communication systems.

\subsection{End-to-End Transceiver Design in IRS-Aided Systems}
For intelligent reflecting surface (IRS) systems, achieving potential advantages depends on accurate CSI, which in practice incurs high computational complexity and feedback overhead. 
Based on the stochastic successive convex approximation (SSCA) algorithm, two DUNNs are designed to jointly unfold the iterative algorithms into a layer-wise structure, to maximize the weighted average sum rate in the IRS system \cite{yanzhen_ris}. The proposed DUNNs consist of a long-term passive beamforming network and a short-term active beamforming network. In particular, the LPBN takes the collected full channel samples as input and outputs the low-dimensional effective CSI, which is processed by SABN to obtain the active beamforming matrices. The proposed DUNNs jointly learn these two beamforming matrices, which reduce the computational complexity and maintain comparable performance. 
Additionally, compared to the SSCA-based optimization approach, the proposed end-to-end DUNNs more closely tie the passive beamforming and the active beamforming matrices.

\section{Open Issues} \label{OpenIssue}

From the above discussion, we can see the great potential of deep-unfolding for transceiver design. Nevertheless, the deep-unfolding is still in its infancy. In this section, we advocate several open issues for future research.

\subsection{Generalized DUNN for Dynamic Environments}
The generalization ability of DUNN has been studied from three aspects in Section \ref{General}. However, the wireless environment generally changes rapidly, such as channel time variation and frequency selectivity, and a wide range of SNR changes etc. Thus, it is difficult to train a DUNN that can adapt to all channel conditions. To further improve the generalization ability, online updating could be a promising technique. It then raises a question: \textit{how to quickly adapt the DUNN model to new scenarios}?
To fast adapt to a new scenario, advanced deep learning technologies of transferring knowledge could be an effective method, including transfer learning and meta learning \cite{meta}.

Transfer learning aims to make full use of prior knowledge in the original scenario when conducting tasks in the new scenario. For example, to verify the generalization ability of DUNN and make it better fit the practical systems, it is trained and tested based on the channel data sampled from the distributions in the original and new scenarios, respectively. Transfer learning can act as a bridge between these two scenarios, making full use of the data in the original scenario to improve the performance of DUNN in the new scenario.

Meta learning aims to find an effective initial model for various new scenarios. Its core idea is to design a fast training method. Generally, a DUNN is trained under certain channel statistics and is limited in a specific scenario, i.e., the performance degrades if the channel statistics change. However, in practical communication systems, there are many scenarios with different channel statistics, such as Rayleigh and Rician distributions, and the channel changes frequently. Training a DUNN that can quickly adapt to various channel statistics is challenging. Meta learning can be employed to find a great initial model that can quickly capture the features of dynamic channels, thus significantly reducing the training overhead.

\subsection{DUNN with Few-Shot Learning}
Although DUNNs achieve great performance and require less training data than the black-box DNNs, training data is scarce and difficult to obtain in many communication scenarios. A promising solution is to employ few-shot learning in DUNN, where the target is to train a powerful DUNN model with very little training data. 

\subsection{Theoretical Analysis} 
The black-box DNNs output unpredictable results and cannot provide theoretical analysis, which restricts its application in practical communication systems. In contrast, the DUNNs are designed by unfolding the iterative algorithms into a layer-wise structure, which is more interpretable than the black-box DNNs. 
However, the trainable parameters in DUNNs are inspired from the structures of black-box DNNs, making the outputs of DUNNs not explainable enough. Therefore, it is important to further improve the interpretability and explainability of DUNNs through theoretical analysis.
Though there have been some preliminary analysis for DUNNs, there still lacks systematic and mathematical foundations for theoretical analysis. The research on establishing the mathematical foundation for the convergence and performance analysis of DUNNs is still in its infancy. In addition, the development of DUNN theory might also inspire more effective training strategies and architectures of DUNNs. 

\subsection{Deployment of Lightweight DUNNs}
In practice, the wireless devices, such as field programmable gate array (FPGA), often have limited storage capacity, computing and communication resources. It might be impractical to assume that all devices could have sufficient capacity to support the deployment of DUNNs. Therefore, developing efficient architectures to balance the performance and deployment cost is an important issue. 
Fortunately, DUNNs have fewer parameters and smaller model size than the black-box DNNs, making it easier to be deployed. To further reduce the deployment cost and make it more affordable, model compression and pruning could be effective means to reduce the network size. 

\subsection{Extension to Promising Communication Scenarios}
As an alternative to conventional algorithms, DUNNs are gaining popularity in wireless communications. Although the idea of deep-unfolding has been applied in many applications, there still exist some communication scenarios where deep-unfolding has not been well investigated. 
The DUNNs can be employed in channel coding to unfold the iterative decoding algorithms, such as successive cancellation decoding of polar codes.
As for CSI feedback, the key idea is to treat the CSI matrix as an image and learn to encode and decode it with a well-designed black-box DNN. The DUNNs can be applied to further compress the CSI matrix, based on its prior knowledge in the existing CSI feedback algorithms.

Moreover, the semantic communication systems are implemented by jointly designing the encoder and decoder modules in conventional communications based on the auto-encoder architecture. DUNNs can be integrated into the joint design of various modules in a semantic communication systems, such as semantic encoder and channel decoder, etc. 
Furthermore, unmanned aerial vehicles (UAVs) equipped with MIMO systems are applied to various communication scenarios, which encourages the hot topic of joint resource allocation and trajectory optimization. The DUNN and other DNNs, e.g., DRL, can be employed to jointly optimize UAVs' trajectory and resource allocation in an end-to-end manner \cite{qiushou_uav}. 

In addition, DUNNs can also be applied to the  emerging MIMO systems, such as space/air/ground MIMO communication systems, Terahertz communication, IRS-aided systems, and integrated sensing and communication (ISAC) systems, etc. In conclusion, the DUNN is a general framework that can be extensively applied to efficiently solve problems in various wireless communication scenarios with satisfactory performance and low complexity. 

\section{Conclusion} \label{Conclusion}
In this article, we provide a comprehensive overview for deep-unfolding enabled transceiver design in next-generation MIMO systems. In particular, we first introduce the general framework of deep-unfolding. We then present some advancements in deep-unfolding, such as adaptive depth and architecture design. Furthermore, some benefits in transceiver technologies for MIMO systems, such as beamforming and channel acquisition, have been illustrated. In addition, the end-to-end DUNNs have been proposed to solve the problems in joint transceiver design. Finally, we highlight some open issues that require further investigation in the future from the perspective of generalization ability, theoretical analysis, and deployment of DUNN, as well as some promising scenarios.

\bibliographystyle{IEEEtran}
\bibliography{IEEEabrv,Reference}

\end{document}